\documentclass[aps,prl,twocolumn,superscriptaddress,noshowpacs]{revtex4}
\usepackage{amsmath}
\usepackage{amssymb}
\usepackage{tikz}
\usepackage{graphicx}
\usepackage{dcolumn}
\usepackage{bm}
\usepackage{epsfig}
\usepackage{psfrag}
\usepackage{latexsym}

\def\II{\hbox{$1\hskip -1.2pt\vrule depth 0pt height 1.6ex width 0.7pt\vrule depth 0pt height 0.3pt width 0.12em$}}
\newcommand{\reffig}[1]{\mbox{Fig.~\ref{#1}}}
\newcommand{\refeq}[1]{\mbox{Eq.~(\ref{#1})}}

\newcommand{\be}{\begin{equation}}
\newcommand{\ee}{\end{equation}}
\newcommand{\ba}{\begin{eqnarray}}
\newcommand{\ea}{\end{eqnarray}}
\renewcommand{\Re}{\mathrm{Re}}
\renewcommand{\Im}{\mathrm{Im}}
\newcommand{\T}{${\mathcal T}\,$}

\begin{document}
\title{Partial time-reversal invariance violation in a flat, superconducting microwave cavity with the shape of a chaotic Africa billiard 
}
\author{B. Dietz}
\altaffiliation{
email:  Dietz@lzu.edu.cn}
\affiliation{%
School of Physical Science and Technology, and Key
Laboratory for Magnetism and Magnetic Materials of MOE, Lanzhou University,
Lanzhou, Gansu 730000, China
}
\author{T. Klaus}
\affiliation{%
  Institut f\"ur Kernphysik,
  Technische Universit\"at Darmstadt,
  D-64289 Darmstadt, Germany
}
\author{M. Miski-Oglu}
\affiliation{%
  GSI Helmholtzzentrum f\"ur Schwerionenforschung GmbH 
  D-64291 Darmstadt, Germany
}
\author{A. Richter}
\altaffiliation{email: Richter@ikp.tu-darmstadt.de }
\affiliation{%
  Institut f\"ur Kernphysik,
  Technische Universit\"at Darmstadt,
  D-64289 Darmstadt, Germany
}
\author{M. Wunderle}
\affiliation{%
  Institut f\"ur Kernphysik,
  Technische Universit\"at Darmstadt,
  D-64289 Darmstadt, Germany
}
\date{\today}
\bigskip
\begin{abstract}
We report on the experimental realization of a flat, superconducting microwave resonator, a microwave billiard, with partially violated time-reversal (\T) invariance, induced by inserting a ferrite into the cavity and magnetizing it with an external magnetic field perpendicular to the resonator plane. In order to prevent its expulsion caused by the Meissner-Ochsenfeld effect we used a cavity of which the top and bottom plate were made from niobium, a superconductor of type II, and cooled it down to liquid-helium temperature $T_{\rm LHe}\simeq 4$~K. The Cavity had the shape of a chaotic Afrivca billiard. Superconductivity rendered possible the accurate determination of complete sequences of the resonance frequencies and of the widths and strengths of the resonances, an indispensable prerequisite for the unambiguous detection of \T invariance violation, especially when it is only partially violated. This allows for the first time the precise specification of the size of \T invariance violation from the fluctuation properties of the resonance frequencies \emph{and} from the strength distribution, which actually depends sensitively on it and thus provides a most suitable measure. For this purpose we derived an analytical expression for the latter which is valid for isolated resonances in the range from no \T invariance violation to complete violation.  
\end{abstract}
\maketitle
{\it Introduction}.--
An important aspect of quantum chaos is the understanding of the features of the classical dynamics in terms of the spectral properties of the corresponding quantum system~\cite{Haake2001,Weidenmueller2009,Gomez2011}. Numerous experimental and numerical studies confirmed that for a fully chaotic classical dynamics they coincide with those of random matrices~\cite{Mehta1990} from the Gaussian orthogonal ensemble (GOE) when time-reversal (\T) invariance is preserved, from the Gaussian unitary ensemble (GUE) when it is violated~\cite{Berry1977,Casati1980,Bohigas1984} and from an ensemble interpolating between the GOE and the GUE, when \T invariance is only partially violated~\cite{Pandey1991,Lenz1992,Hul2004}. Most suitable for the experimental verification are flat, cylindrical microwave resonators~\cite{StoeckmannBuch2000,Richter1999} and microwave networks~\cite{Hul2004}. For microwave frequencies below a certain cutoff value $f_{\rm max}$, the associated Helmholtz equation is mathematically equivalent to the Schr\"odinger equation of the quantum billiard and the quantum graph of corresponding shape~\cite{Kottos1997,Kottos1999}, respectively. The random-matrix theory (RMT) analysis of the spectral properties of a quantum system and the assignment to one of these ensembles requires complete sequences of several hundreds of eigenvalues~\cite{Bohigas1984} or an elaborate procedure to cope with missing levels~\cite{Liou1972,Zimmermann1988,Agvaanluvsan2003,Bohigas2004,Frisch2014,Mur2015,Bialous2016} which hinder or render the unambiguous determination of the strength of \T invariance violation unfeasible in cases where it is only partially violated. For \T invariant systems complete sequences of up to 5000 eigenvalues~\cite{Dietz2015b,Dietz2015,Dietz2018} of the corresponding quantum billiard were obtained in high-precision experiments at liquid-helium temperature $T_{\rm LHe}=4$~K with niobium and lead-coated microwave resonators which become superconducting at $T_c=9.2$~K and $T_c=7.2$~K, respectively. Quantum systems in the presence \T invariance violation were investigated experimentally, e.g., in nuclear spectra and reactions~\cite{French1985,mitchell2010}, through Rydberg excitons~\cite{Assmann2016} in copper oxide crystals, and in electron transport through quantum dots~\cite{Pluhar1995}. The effects of \T invariance violation on the fluctuation properties in the eigenvalue spectra have been measured in microwave billiards~\cite{So1995,Stoffregen1995,Wu1998} and networks~\cite{Hul2012,Bialous2016,Allgaier2014}. However, one had to cope with missing levels in these experiments. Furthermore, the fluctuation properties of the scattering matrix describing open quantum systems with partially violated \T invariance were studied thoroughly with microwave billiards~\cite{Dietz2007a,Dietz2009a,Dietz2010}. In this letter we present the first experimental realization of \T invariance violation in a superconducting microwave billiard, which allows the precise determination of the strengths and widths of the resonances and of their positions, i.e., of the resonance frequencies and, thereby, of the size of \T invariance violation based on the strength distribution and the spectral fluctuation properties.  

In the room-temperature experiments \T invariance violation was induced by inserting a ferrite into the resonator and magnetizing it with an external magnetic field. Due to the Meissner-Ochsenfeld effect~\cite{Meissner1933} this, however, is no longer possible in experiments with superconducting lead-coated cavities~\cite{Dietz2015}, since lead is a superconductor of type I~\cite{Onnes1911} which is characterized by zero electrical resistance and perfect diamagnetism. Meissner and Ochsenfeld showed that regardless of whether such a superconductor was cooled below the critical temperature and placed in a magnetic field or placed in a magnetic field and then cooled below $\rm T_c$ the field was expelled from the superconductor. Superconductors of type I can be turned into normal conductors by increasing the temperature or the external magnetic field beyond their critical values $\rm T_c$ and $B_{c}$, respectively, whereas niobium is one of type II~\cite{Shubnikov1937} for magnetic fields between two critical values, $B_{\rm c1}=153$~mT and $B_{\rm c2}=268$~mT. Below $B_{c1}$ it behaves like a superconductor of type I and above $B_{\rm c2}$ it becomes normal conducting. Between these critical values, magnetic flux starts to penetrate the superconductor via vortices corresponding to regions of circulating supercurrent around a core which essentially behaves like a normal conductor~\cite{Abrikosov1957}. The magnetic flux enters the superconductor through these flux tubes, with the supercurrents screening the bulk material from the external field. We used this property of niobium to achieve partial \T invariance violation at $T_{\rm LHe}$.

\begin{figure}[h!]
\includegraphics[width=0.7\linewidth]{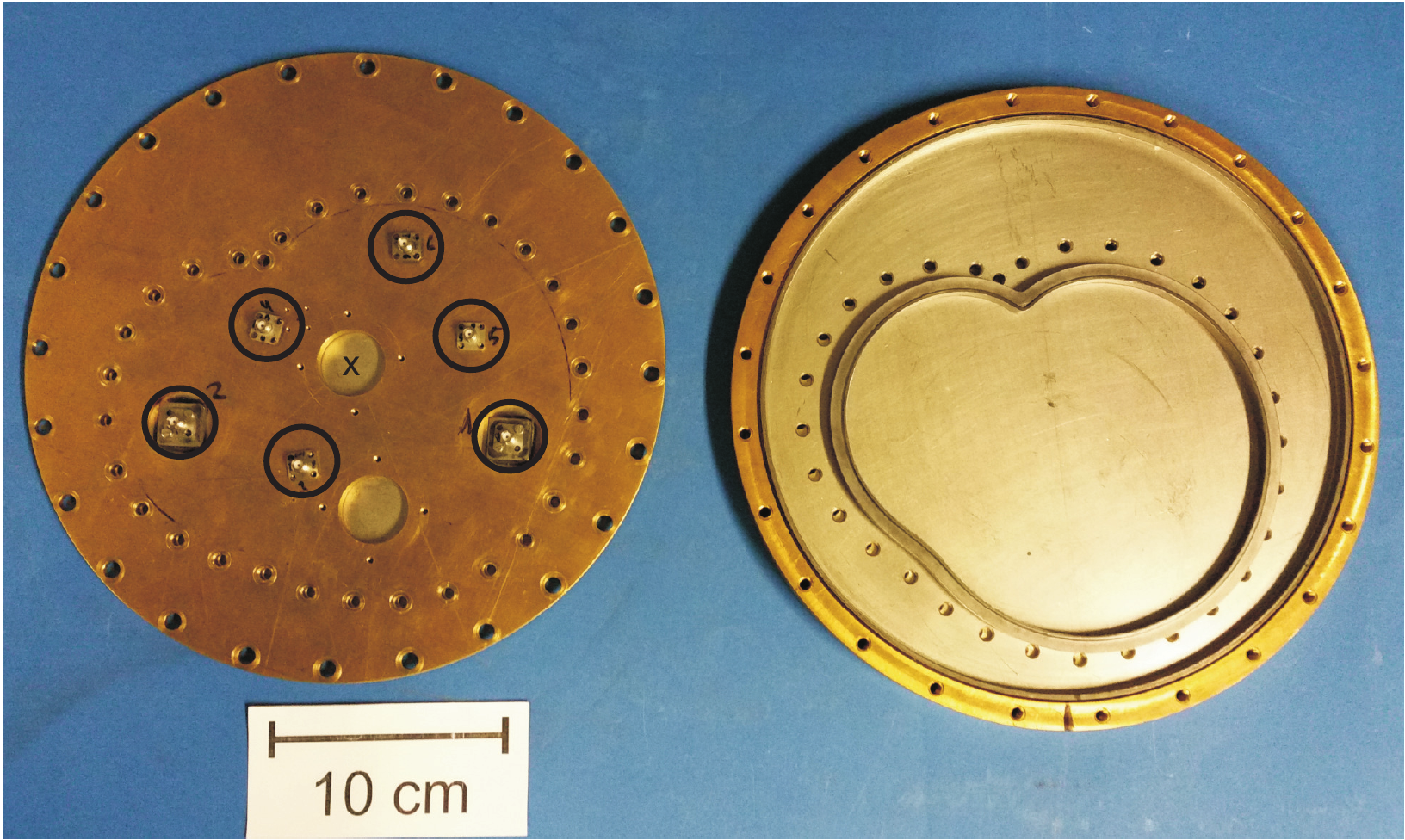}
	\caption{Photograph of the top plate (left) and basin (right) of the cavity. The circles on the top plate indicate the six different positions where two antennas were attached and the cross marks the ferrite position. A lead-coated brass frame with the shape of an Africa billiard was inserted into the circular basin which was made from niobium.
}
\label{fig1}
\end{figure}

{\it Experiment}.--
The microwave billiard consisted of a circular niobium billiard of 7~mm height, previously used for the experimental investigation of chaos-assisted tunneling~\cite{Dembowski2000a}, containing a brass frame of 7.1~mm height with the shape of a fully chaotic Africa billiard, see~\reffig{fig1}, which was coated with lead. In order to guarantee a good electrical contact between the frame and the top and bottom plates, the upper and lower edges of the frame were sharp cut and the resonator was inserted into a circular brass container which was flush with the former and tightly closed with screws, thus firmly pressing the top and bottom niobium plates onto the frame. The screw holes are visible in~\reffig{fig1}. In order to induce \T invariance violation, a cylindrical CV19 ferrite of 5~mm height and 4~mm diameter was introduced into the billiard at the position marked by a black cross in~\reffig{fig1} and magnetized by an external magnetic field perpendicular to the plates generated by two NdFeB magnets mounted outside the resonator above and below the ferrite at a distance of 3.5~mm from the Niobium plates, which corresponds to a magnetic field strength of $B=200$~mT~\cite{Dietz2009a,Dietz2010}, such that $B_{c_1}<B<B_{c_2}$. 

\begin{figure}[t!]
	\includegraphics[width=0.9\linewidth]{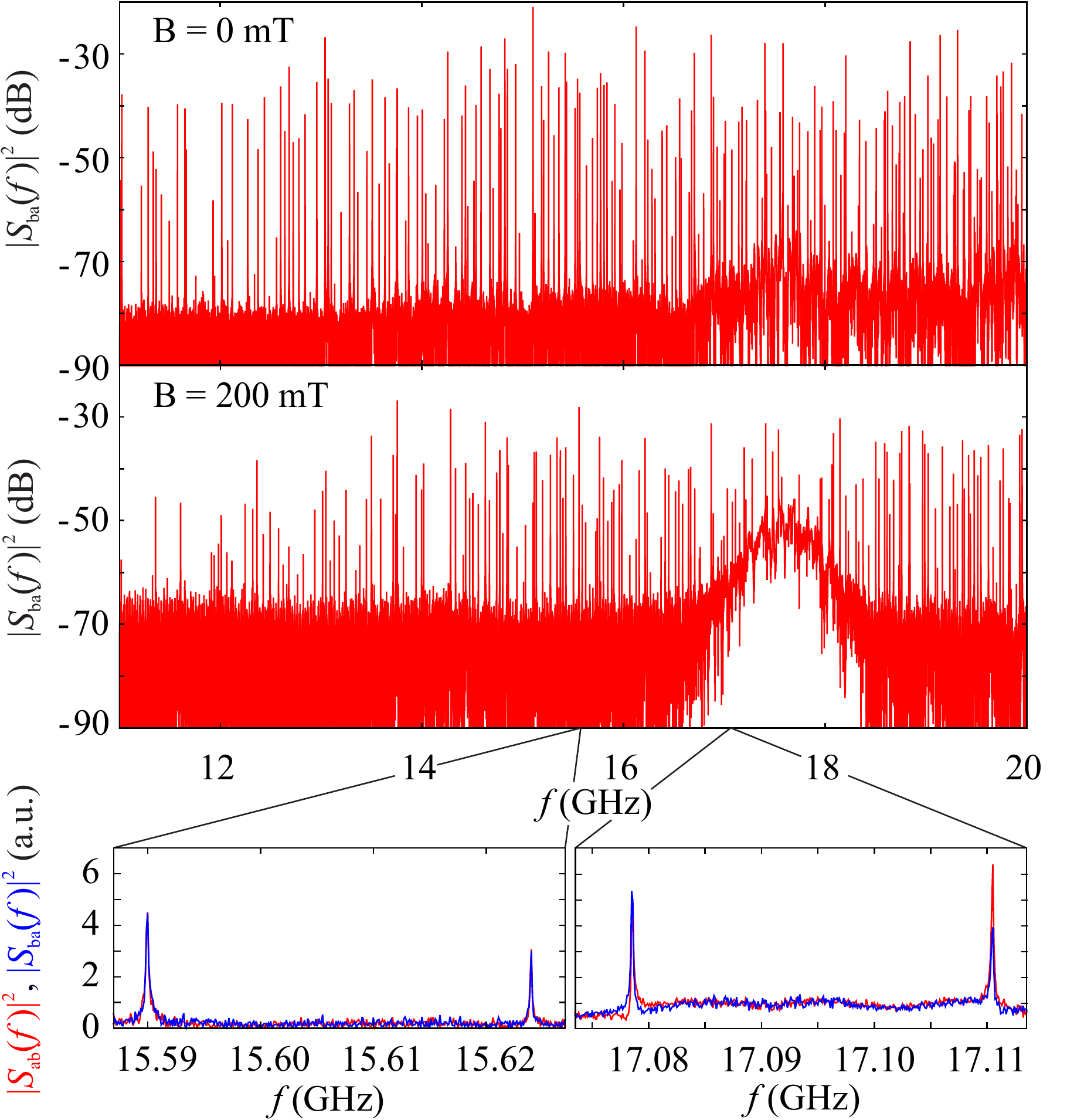}
	\caption{Transmission spectra for $B=0$~mT (top) and $B=200$~mT (middle). The latter exhibits a broad peak in the frequency range $f\in [16.5,18.5]$~GHz, which is attributed to trapped modes inside the ferrite. The zooms into this spectrum (bottom) illustrate exemplary the conservation (left) and violation (right) of the principle of reciprocity below and within this frequency range.
}
\label{fig2}
\end{figure}

For the measurement of the reflection and transmission spectra two antennas were attached at two out of six possible positions visible in~\reffig{fig1} and connected to a Vector Network Analyzer (VNA) which provided the rf signal for frequencies below $f_{\rm max}=20$~GHz. The VNA determined the amplitude and phase of the output signal relative to the input signal, thus yielding the scattering ($S$) matrix elements $S_{\rm ba}$ describing the scattering process from antenna $a$ to $b$. Figure~\ref{fig2} shows transmission spectra measured at $B=0$~mT and $B=200$~mT, respectively. The latter one exhibits a broad peak in the frequency range [16.5,18.5]~GHz which may be attributed to electric field modes trapped inside the ferrite (see below). While for $B=0$~mT and also below and above that range for $B=200$~mT the principle of reciprocity holds, i.e., $S_{\rm ab}(f)=S_{\rm ba}(f)$, it is clearly violated within it. This is illustrated in the two zooms shown at the bottom of~\reffig{fig2}.

{\it Fluctuations in the transmission spectra}.--
We use the scattering matrix approach~\cite{Mahaux1969} for the RMT description of the fluctuation properties in the transmission spectra. It was developed by Mahaux and Weidenm{\"u}ller in the context of compound nuclear reactions and extended to microwave resonators in~\cite{Albeverio1996},
\begin{equation}
	S_{\rm ba}(f) = \delta_{\rm ba} - 2\pi i\left[\hat W^\dagger\left(f\II-\hat H^{eff}\right)^{-1}\hat W\right]_{\rm ba}.
\label{eqn:Sab}
\end{equation}
Here, $\hat H^{eff}=\hat H-i\pi\hat W\hat W^\dagger$ with $\hat H$ denoting the Hamiltonian describing the closed resonator or quantum billiard and $\hat W$ accounting for the coupling of the resonator modes to their environment. Since the classical dynamics of the Africa billiard is chaotic, we model $\hat H$ by an ensemble of $N\times N$-dimensional random matrices with entries
\begin{equation}
        H_{\rm\mu\nu}=H_{\rm\mu\nu}^{(S)}+i\frac{\pi\xi}{\sqrt{N}}
        H_{\rm\mu\nu}^{(A)}.
\label{eqn:hamiltonian}
\end{equation}
Here, $\hat H^{(S)}$ is a real-symmetric random matrix from the GOE and $\hat H^{(A)}$ is a real-antisymmetric one with uncorrelated Gaussian-distributed matrix elements. The parameter $\xi$ determines the magnitude of \T violation. For $\xi =0$ $\hat H$ describes chaotic systems with preserved \T invariance, whereas for $\pi \xi / \sqrt{N} = 1$ $\hat H$ is a random matrix from the GUE. Yet, the transition from GOE to GUE already takes place for $\xi\simeq 1$~\cite{Dietz2010}. 

The entries of $\hat W$ were real, Gaussian distributed with $W_{\rm a \mu}$ and $W_{\rm b \mu}$ describing the coupling of the antenna channels to the resonator modes. Furthermore, $\Lambda$ equal fictituous channels accounted for the Ohmic losses in the ferrite~\cite{Dietz2007a} and the walls of the resonator~\cite{Dietz2009a,Dietz2010}. Direct transmission between the antennas was negligible, that is, the frequency-averaged $S$-matrix was diagonal. This feature is incorporated in the RMT model through the property $\sum_{\mu = 1}^N W_{\rm e \mu} W_{\rm e^\prime \mu}=N v_{\rm e}^2 \delta_{\rm ee^\prime}$. The parameter $v^2_{\rm e}$ measures the average strength of the coupling of the resonances to channel $e$, which for $e=a,\ b$ is determined by the average size of the electric field at the position of the antenna. Generally, it is related to the transmission coefficients $T_{\rm e} = 1 - |\left\langle{S_{\rm ee}}\right\rangle |^2$ via $T_{\rm e} = \frac{4 \pi^2 v^2_{\rm e} / d}{(1 + \pi^2 v^2_{\rm e} / d)^2}$ with $d=\sqrt{\frac{2}{N}\langle H_{\rm\mu\mu}^2\rangle}\frac{\pi}{N}$ denoting the mean resonance spacing. The input parameters of the RMT model~\refeq{eqn:Sab} are the transmission coefficients $T_{\rm a}, T_{\rm b}$ associated with antennas $a$ and $b$, $\tau_{\rm abs}=\Lambda T_{\rm c}$ accounting for the Ohmic losses, and the \T violation parameter $\xi$. In the RMT model~\refeq{eqn:Sab} the coupling matrix $\hat W$ is assumed to be frequency independent. Accordingly, in order to ensure this for the resonance widths, the frequency range used for the analysis of the data was divided into windows of 1~GHz as in the room-temperature measurements~\cite{Dietz2010}. In~\cite{Dietz2009a,Dietz2010} the parameters $\xi$, $T_{\rm a,b}$ and $\tau_{\rm abs}$ were deduced from the cross-correlation coefficients and two-point $S$-matrix correlation functions. However, since for well-isolated resonances the distributions of the resonance strengths and $S$-matrix elements depend more sensitively on  $\xi$, $T_{\rm a,b}$ and $\tau_{\rm abs}$, we used them. They were obtained by averaging over six transmission spectra. 

The \T invariance violation parameter $\xi$ was determined by fitting an analytic expression for the distribution of the strengths $y_{\rm ab}=\Gamma_{\rm \mu a}\Gamma_{\rm\mu b}$ to the experimentally determined one. Here, $\Gamma_{\rm \mu a}$ and $\Gamma_{\rm \mu b}$ are the partial widths related to antennas $a,\, b$, which are proportional to the electric field intensity at the position of the antenna. For sufficiently isolated resonances the $S$-matrix has the form 
\begin{equation}
S_{\rm a b} = \delta_{\rm a b} 
    - i\frac{\sqrt{\Gamma_{\rm \mu a}\Gamma_{\rm\mu b}}}
	     {f-f_{\rm\mu} + \frac{i}{2}\Gamma_{\rm\mu}
            },
\label{SMatrix}
\end{equation}
close to the $\mu$th eigenfrequency $f_{\mu}$ with $\Gamma_{\rm\mu}$ denoting the total width of the corresponding resonance~\cite{Alt1995}. Since this is the case for transmission spectra obtained from measurements at superconducting conditions, we may determine the strength of the resonances with high precision by fitting this expression to them. Note that the individual partial widths are not accessible~\cite{Dembowski2005}. 

The strength distribution is derived from that of the partial widths~\cite{Dembowski2005,Dietz2006a}, $P(t_{\rm a})$ with $t_{\rm a}=\Gamma_{\rm \mu a}$, which for chaotic systems with preserved \T invariance ($\xi =0$) is a Porter-Thomas distribution, whereas it is an exponential one in the case of complete \T violation~\cite{Guhr1998}. For the case of partial \T violation it is obtained by starting from $H_{\mu\nu}= \Re H_{\mu\nu} +i\lambda \Im H_{\mu\nu}$ with $\lambda =\frac{\pi\xi}{\sqrt{N}}$ and proceeding as for the GUE, which in the limit of large $N$~\cite{Guhr1998} amounts to computing the distribution of $\left(\Re H_{\mu\nu}\right)^2+\lambda^2\left(\Im H_{\mu\nu}\right)^2$, $P(t_{\rm a})=\frac{1}{2\pi}\int_{-\infty}^\infty {\rm d}xe^{-x^2/2}\int_{-\infty}^\infty {\rm d}y e^{-y^2/2}\delta\left(t_{\rm a}-\sqrt{x^2+\lambda^2y^2}\right)$, yielding
\be
P(t_{\rm a})=\frac{1}{\lambda}e^{-\frac{1+\lambda^2}{2\lambda^2}t_{\rm a}}I_0\left(\frac{1-\lambda^2}{2\lambda^2}t_{\rm a}\right),
\label{Anal}
\ee
where $I_0(x)$ is the zero-order modified Bessel function of the first kind. Here, $P(t_{\rm a})$ is normalized to unity and its first moment is $\tau_a=\frac{1+\lambda^2}{2}$. 

\begin{figure}[h!]
\includegraphics[width=0.8\linewidth]{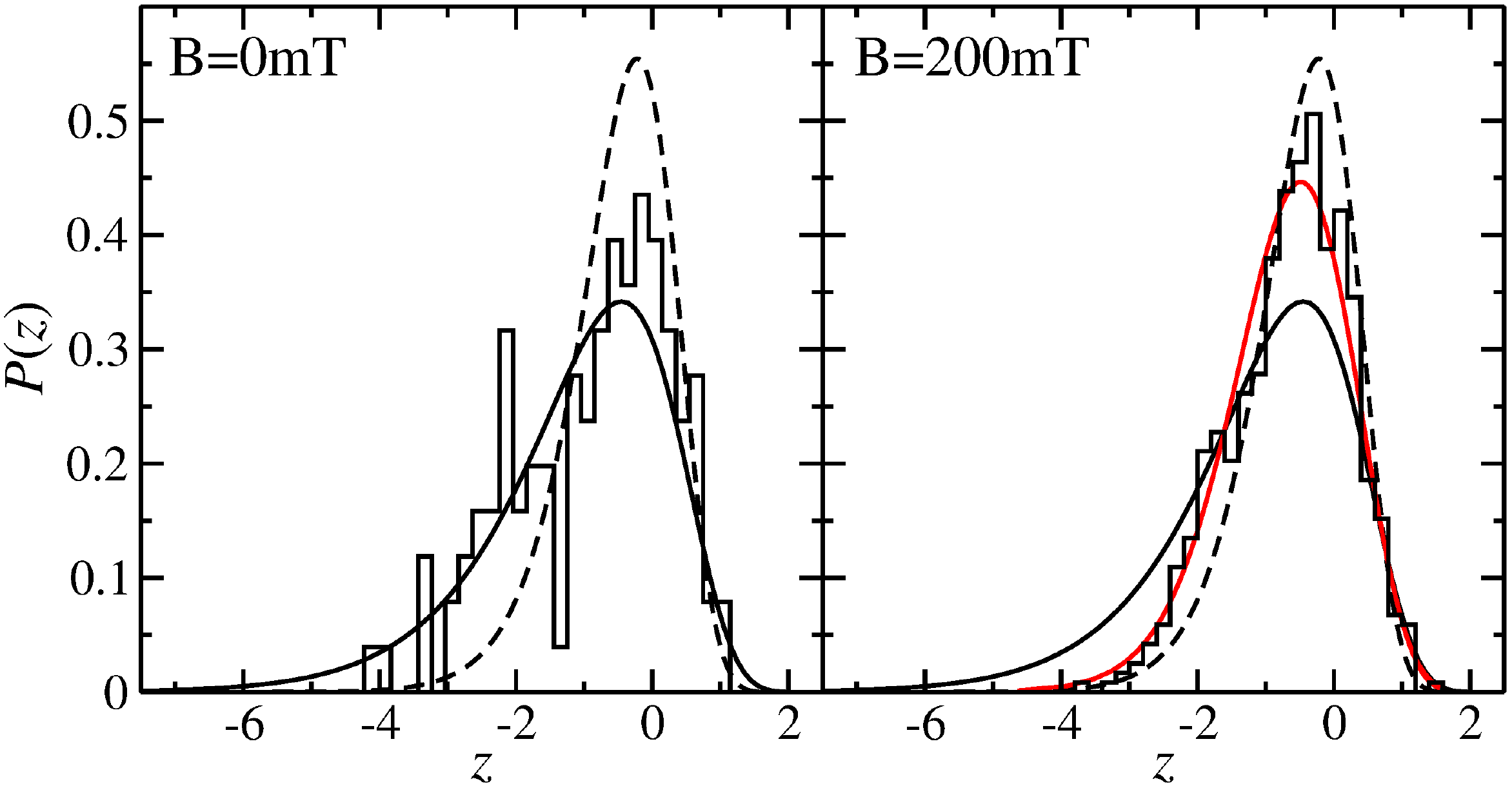}
	\caption{Experimental strength distribution (histogram) for external magnetic fields $B=0$~mT (left) and $B=200$~mT (right) in comparison to the GOE (solid lines) and GUE (dashed lines) curves. The red curve shows the best fitting analytical curve obtained from~\refeq{Anal}, yielding $\xi=0.2$
}
\label{fig4}
\end{figure}

We proceeded as in Ref.~\cite{Dembowski2005} to determine the experimental distributions. The $N_{\rm ab}$ identified products of partial widths were rescaled such that the expectation value of the strength distribution, $\tau_{\rm a}\tau_{\rm b} =N_{\rm ab}^{-1}\sum_{\mu =1}^{N_{\rm ab}}\Gamma_{\rm\mu a}\Gamma_{\rm\mu b}$, equaled unity. Figure~\ref{fig4} shows the strength distribution for the GOE (solid line), the GUE (dashed line) and the experimental results for $B=0$~mT (left panel) and $B=200$~mT (right panel) on a logarithmic scale, $z=\log_{10}\left(\frac{y}{\tau_{\rm a}\tau_{\rm b}}\right)$~\cite{Adams1998}. The red solid line shows the analytical strength distribution deduced from~\refeq{Anal} which best fits the experimental one for $\xi =0.2$. Actually, this procedure for determining $\xi$ is less demanding than those based on the $S$-matrix correlation functions, since the analytical expression is simpler and there are no ambiguities to overcome as encountered in Refs.~\cite{Dietz2009a,Dietz2010}.

\begin{figure}[h!]
\includegraphics[width=0.6\linewidth]{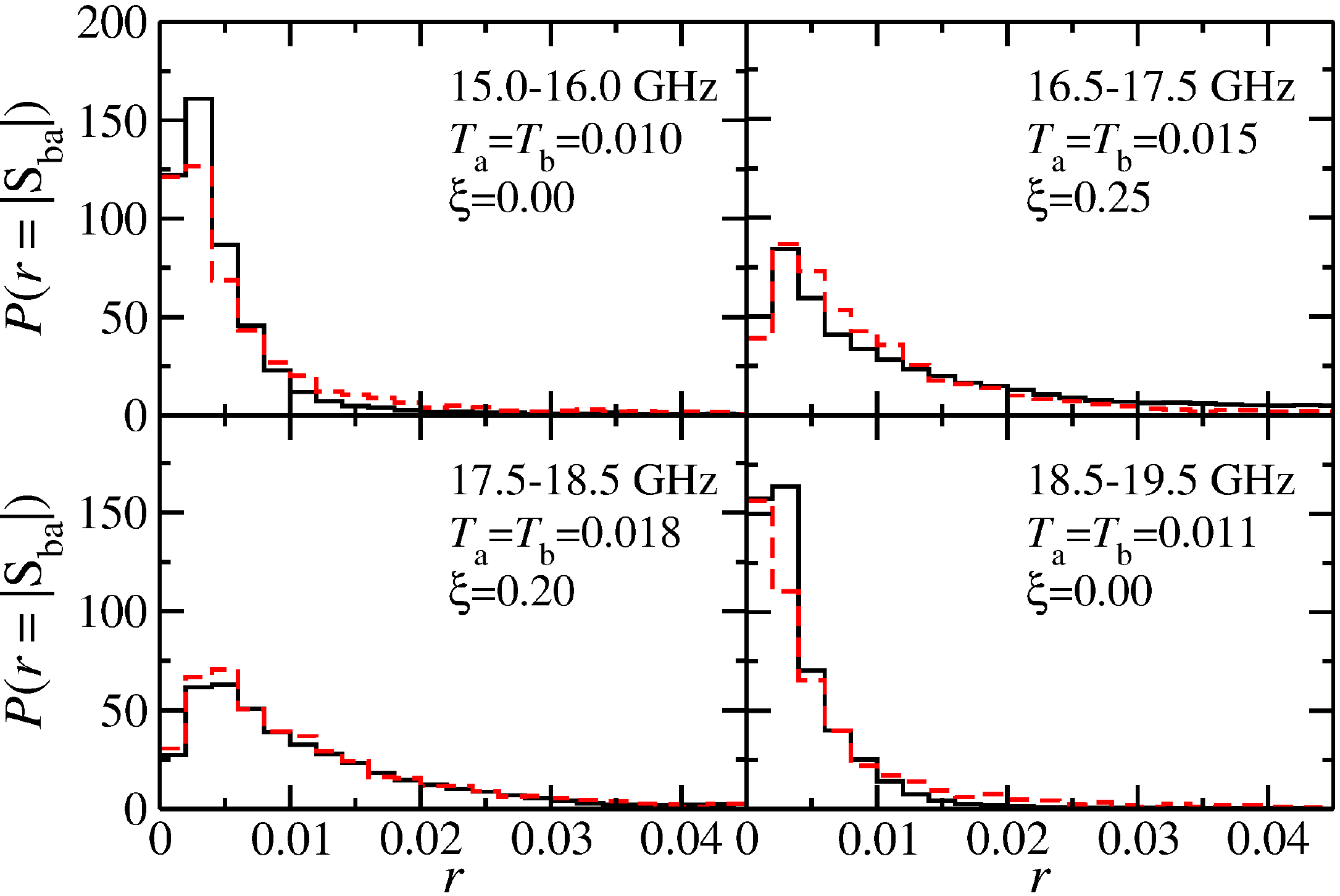}
	\caption{Distribution of the modulus of the measured off-diagonal $S$-matrix element $\vert S_{\rm ba}\vert$ (black solid lines) in comparison to RMT simulations (red dashed lines). The frequency ranges and values of the transmission coefficients are provided in the insets.
}
\label{fig3}
\end{figure}

In order to obtain $T_{\rm a,b}$ and $\tau_{\rm abs}$ we generated ensembles of 500 random $S$ matrices of the form \refeq{eqn:Sab} with $\Lambda=30$ and $N=200$, and fit the resulting $S$-matrix distributions to the experimental ones. Since they depend sensitively on these parameters their values could be determined with a high accuracy. They are given in the insets of ~\reffig{fig3} which exhibits the distribution of the $S$-matrix element $S_{\rm ba}$ in the frequency intervals $[15,16]$, $[16.5,17.5]$, $[17.5,18.5]$ and $[18.5,19.5]$~GHz. Generally, absorption was very small in comparison to that in the room temperature experiments~\cite{Dietz2009a,Dietz2010}. Indeed, the experimental distributions are well described for a fixed $\tau_{\rm abs}=0.0001$. The distributions clearly reflect the distinct features in the transmission spectrum observed in~\reffig{fig2} for $B=200$~mT within and outside the range $16.5-18.5$~GHz. Interestingly, below and above this frequency range $T_{\rm a}$ and $T_{\rm b}$ barely vary whereas they are considerably larger within it, even though they are expected to increase slowly with frequency~\cite{Dietz2010}. This indicates that there the electric field distribution, and therefore that at the antennas, is noticeably influenced by the presence of the magnetized ferrite. This issue will be further discussed below. 

\begin{figure}[h!]
\includegraphics[width=0.49\linewidth]{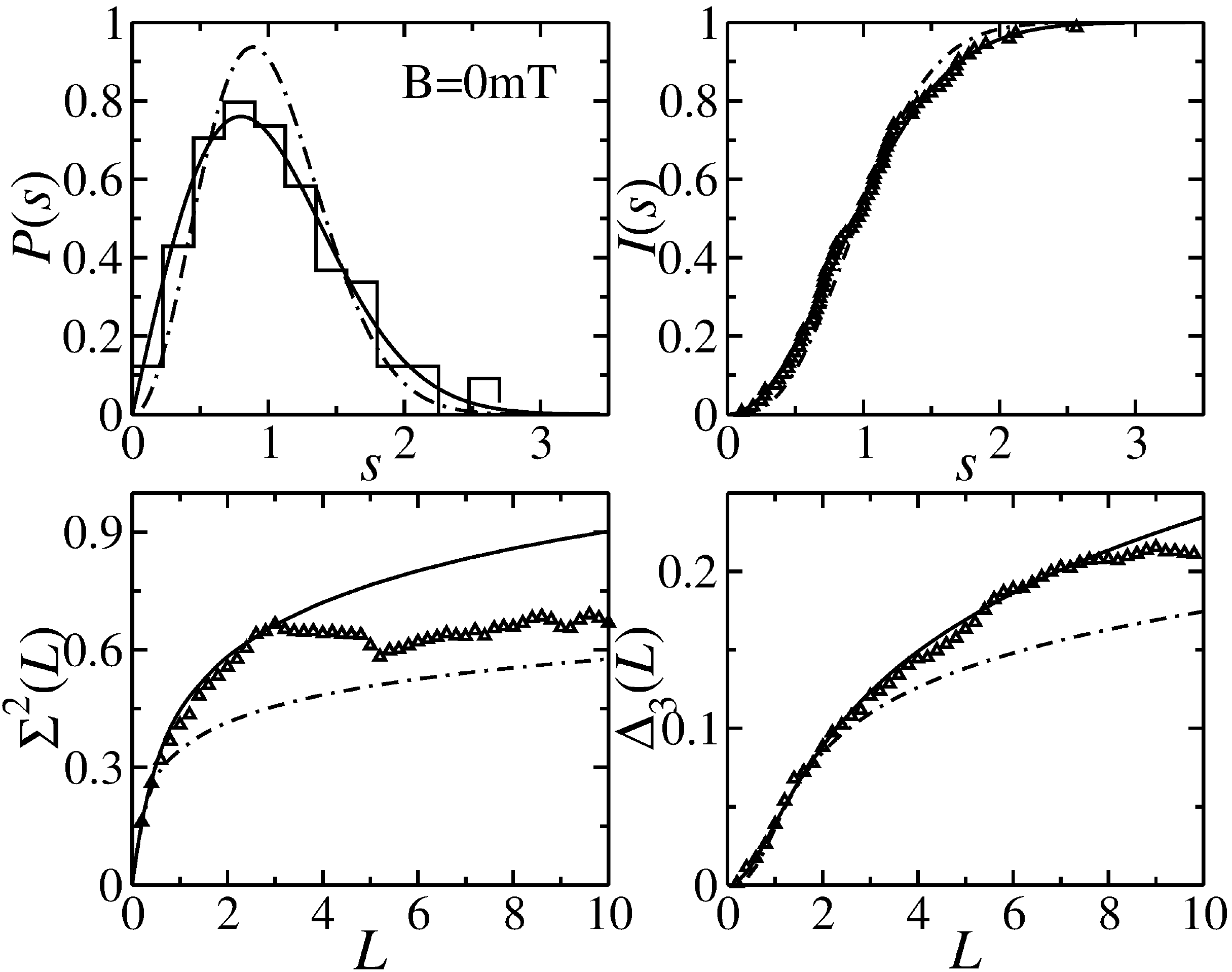}
\includegraphics[width=0.49\linewidth]{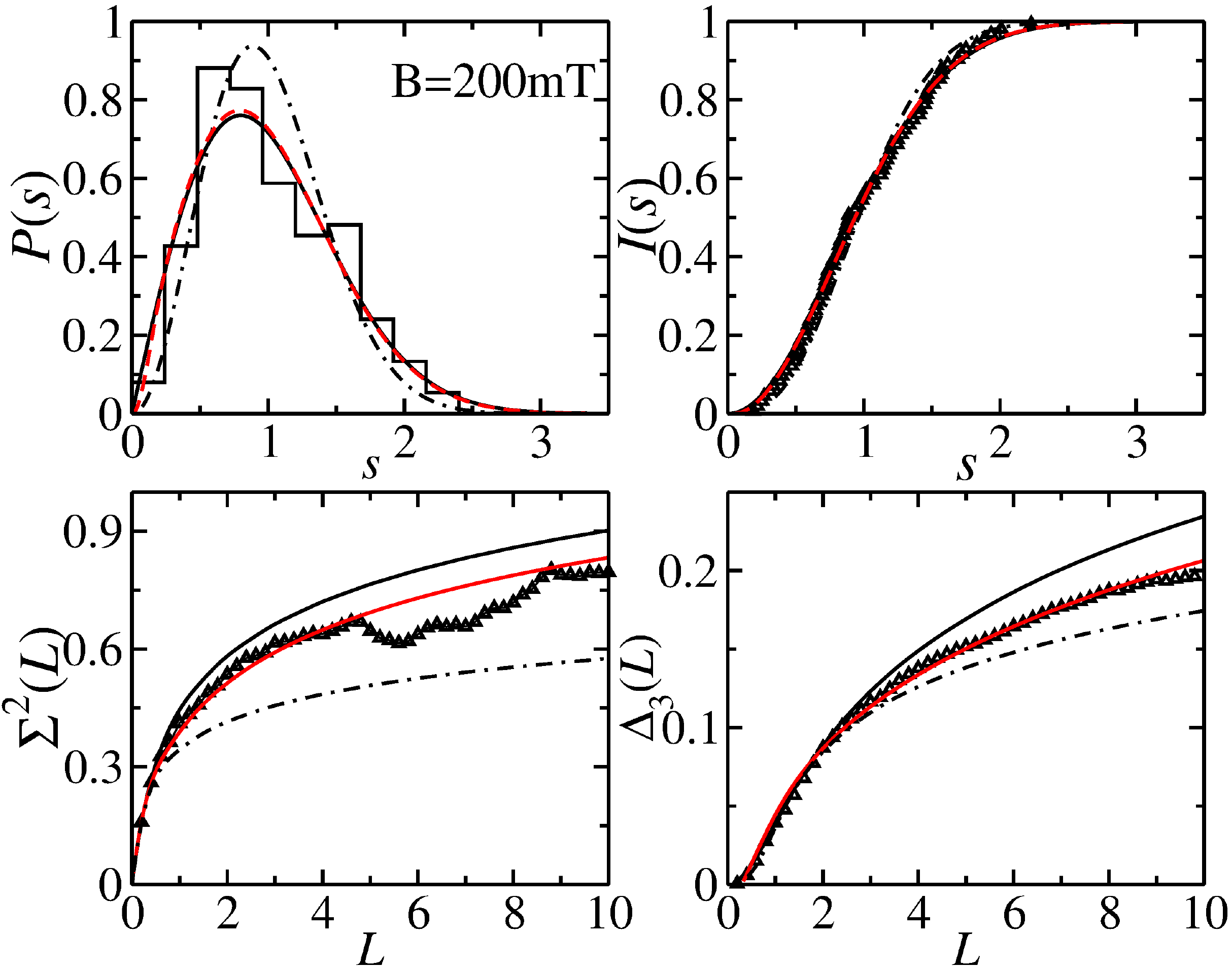}
	\caption{ Spectral properties (histograms and triangles) for external magnetic fields $B=0$~mT (left) and $B=200$~mT (right). The red solid lines show the corresponding best fitting curves computed from Eqs.~(\ref{ps}) and~(\ref{yl}), yielding $\xi=0.2$.
}
\label{fig5}
\end{figure}

{\it Fluctuation properties of the resonance frequencies}.--
In order to verify the value of $\xi$ we, furthermore, investigated the spectral properties of the Africa billiard without and with magnetized ferrite. Due to the small area of the Africa microwave billiard, the complete sequences of resonance frequencies comprised in both cases only $\approx 215$ levels in the interval $[12,19]$~GHz. The resonance frequencies $f_\mu$ were unfolded to mean spacing unity~\cite{StoeckmannBuch2000,Haake2001} with Weyl's law, $\epsilon_\mu =\frac{A\pi}{c_0^2}f_\mu^2+\frac{\mathcal{L}}{2c_0}f_\mu+$ const., where $A\simeq 230$~cm$^2$, $\mathcal{L}\simeq 53$~cm and $c_0$ denote the area, perimeter, and speed of light, respectively. We considered the distribution $P(s)$ of nearest-neighbor spacings $s_{\rm i}=\epsilon_{\rm i+1}-\epsilon_{\rm i}$, the cumulative nearest-neighbor spacing distribution $I(s)$, the variance $\Sigma^2(L)=\langle\left(N(L) -L\right)^2\rangle$ of the number of levels in an interval $L$, where $\langle N(L)\rangle =L$, and the Dyson-Mehta statistic $\Delta_3(L)$ which gives the average least-square deviation of $N(L)$ from a straight line~\cite{Mehta1990}. These measures have the advantage that analytical expressions exist for the transition from GOE to GUE. The nearest-neighbor spacing distribution is given by~\cite{Lenz1992}
\be
P(s;\xi)=\sqrt{\frac{2+\xi^2}{2}}sc^2(\xi){\rm erf}\left(\frac{sc(\xi)}{\xi}\right)e^{-\frac{s^2c(\xi)^2}{2}}
\label{ps}
\ee
with $c(\xi)=\sqrt{\pi\frac{2+\xi^2}{4}}\left[1-\frac{2}{\pi}\left(\tan^{-1}\left(\frac{\xi}{\sqrt{2}}\right)-\frac{\sqrt{2}\xi}{2+\xi^2}\right)\right]$ and ${\rm erf}(x)$ denoting the error function. To compute $\Sigma^2(L;\xi)$ and $\Delta_3(L;\xi)$ we used analytical results for the two-point cluster function~\cite{Mehta1990},
\begin{equation}
Y_2(L;\xi)=\det\left[
\begin{pmatrix}
s(L) &-D(L;\xi)\\
-J(L;\xi) &s(L)
\end{pmatrix}
\right],
\label{yl}
\end{equation}
with $s(L)=\frac{\sin\pi L}{\pi L}$, $D(L;\xi)=\frac{1}{\pi}\int_0^\pi{\rm d}xe^{2\xi^2 x^2}x\sin(Lx)$ and $J(L;\xi)=\frac{1}{\pi}\int_\pi^\infty{\rm d}xe^{-2\xi^2x^2}\frac{\sin(Lx)}{x}$~\cite{Pandey1991,Bohigas1995}.
We determined the \T invariance violation parameter $\xi$ by fitting these analytical curves to the corresponding experimental ones, yielding for $\xi$ the same value as for the strength distribution. In~\reffig{fig5} we compare the experimental curves (histograms and triangles) for the microwave billiard without (left) and with magnetized ferrite (right) with those for the GOE (solid lines) and the GUE (dashed lines). Furthermore, the red solid lines show the corresponding analytical curves for $\xi =0.2$. Note, that they barely differ from the GOE curves for $P(s)$ and $I(s)$, that is, it is indispensable to consider in addition long-range correlations to detect the partial \T violation. 

{\it Conclusions}.--
We may conclude from our findings, that \T invariance is partially violated in the frequency range from 16.5-18.5~GHz. Yet, we demonstrated in~\cite{Dietz2007a} that this may be achieved only if the resonances overlap~\cite{Dietz2009a,Dietz2010}, whereas in the experiments discussed in the present letter they were isolated even in the frequency range, where \T violation was observed. Indeed, violation of the principle of reciprocity is induced if the coupling of the spins in the ferrite -- precessing with their Larmor frequency about the external magnetic field -- to the rf magnetic-field components of the resonator modes depends on the rotational direction of polarization of the latter, implying that the modes should be circularly polarized with unequal magnitudes of the two rotational components~\cite{Dietz2007a}. Furthermore, the effect is strongest at the ferromagnetic resonance which is at around 6~GHz. However, in both the experiments at room temperature and at 4~K, a stronger violation of \T invariance was observed at about 15 and 23~GHz and around 17.5~GHz, respectively. This finding was attributed to modes trapped inside the ferrite in these frequency regions. To confirm this assumption we performed simulations with CST MICROWAVE STUDIO~\cite{CST}. Indeed, since the dielectric constant of the cylindrically-shaped ferrite is larger than that of air, TE field modes of Bessel-function ($J_m(x),\, m=0,1,2,\cdots$) type may be localized inside the ferrite at its resonance frequencies~\cite{Hentschel2002b}. In order to find them, we computed the ratio of the electric energy stored inside the ferrite and the resonator, respectively. The magnitudes of the two rotational components of the circularly polarized microwaves excited inside the resonator become unequal through the coupling to trapped modes with $m>0$, which is possible only if the electric field intensity is non-vanishing in the vicinity of the ferrite. The numerical simulations revealed that this indeed is the case for trapped modes identified around 15~GHz and 23~GHz thus confirming the interpretation of the results obtained in the room-temperature experiments~\cite{Dietz2010}. A crucial difference between the present and previous experiments is that the height of the cavity was larger in the latter case so that there was a gap between the ferrite and the top plate. This was taken into account in further simulations which clearly showed, that there is one trapped mode at about 18.3~GHz, which leads to the broad peak in the transmission spectrum in~\reffig{fig2}, and interacts with the resonator modes. Thus, the numerical simulations firstly confirm our interpretation of the mechanism which leads to partial violation of \T invariance and furthermore provide an explanation of our finding that it occurs around 18 GHz. Summarizing, we observe only partial \T invariance violation, and this only if, in the region of the ferromagnetic resonance, the electric field intensity, i.e., the modulus of the wavefunction, is nonvanishing at the position of the ferrite. This fact may be decribed in line with the approach presented for charged particles in~\cite{Bohigas1995} in a ray-dynamical picture with the effect of the magnetized ferrite on the microwaves modeled by an average potential of finite range. Namely, only those trajectories -- which, in distinction to Ref.~\cite{Bohigas1995} are all straight in our case -- are influenced which pass the ferrite within this range. Accordingly, the strength of \T invariance violation $\xi$ is proportional to the ratio of the effective area occupied by the potential and the area of the billiard. Within this picture only part of the resonances will be influenced by the magnetized ferrite, as indeed observed in the experiments. Furthermore, it is corroborated by our result that the value of $\xi$ is the same in the experiments with normal and superconducting microwave billiards, respectively. This explains why we observe only partial \T invariance violation but, since the details of the interaction are unknown, we cannot provide a quantitative prediction for $\xi$ as in Ref.~\cite{Bohigas1995}.     

In summary, we experimentally realized for the first time a superconducting microwave billiard with partially violated \T invariance. In order to obtain longer and complete sequences of eigenvalues and to realize a stronger violation of \T invariance the area of the billiard needs to be increased and a few more ferrites must be added. Yet, care has to be taken that additional absorption leaves the resonances isolated in order to ensure complete level sequences.

\begin{acknowledgments}
This work was supported by the Deutsche Forschungsgemeinschaft (DFG) within the Collaborative Research Center 1245. BD thanks the NSF of China for financial support under Grant Nos. 11775100 and 11961131009.
\end{acknowledgments}
\newpage
\bibliography{References}
\end{document}